# Megabits secure key rate quantum key distribution

Qiang Zhang[1,2], Hiroki Takesue[3], Toshimori Honjo[3], Kai Wen[1], Toru Hirohata[4], Motohiro Suyama[5], Yoshihiro Takiguchi[4], Hidehiko Kamada[3], Yasuhiro Tokura[3], Osamu Tadanaga[3], Yoshiki Nishida[3], Masaki Asobe[3], Yoshihisa Yamamoto[1,2]

[1] Edward L. Ginzton Laboratory, Stanford University, Stanford, California 94305, USA
[2] National Institute of Informatics, 2-1-2 Hitotsubashi, Chiyoda-ku, Tokyo, 101-843, Japan
[3] NTT Basic Research Laboratories, NTT Corporation, 3-1 Morinosato Wakamiya, Atsugi, Kanagawa 243-0198, Japan
[4] Central Research Laboratory, Hamamatsu Photonics K.K.
[5] Electron Tube Division, Hamamatsu Photonics K.K.

**Quantum cryptography (QC)[1] can provide unconditional secure communication between two authorized parties based on the basic principles of quantum mechanics[2-4]. However, imperfect practical conditions limit its transmission distance and communication speed. Here we implemented the differential phase shift[5] (DPS) quantum key distribution (QKD) with up-conversion assisted hybrid photon detector[6] (HPD) and achieved 1.3 M bits per second secure key rate over a 10-km fiber, which is tolerant against the photon number splitting (PNS) attack[7,8], general collective attacks on individual photons[9], and any other known sequential unambiguous state discrimination (USD) attacks[10,11].**

QC over a fiber link is thought to be the first possible practical applications in quantum information research[1]. Currently high bit rate QKD has attracted more and more effort[12-19]. The main limited ingredient for high speed QKD is the imperfection of single photon sources and single photon detectors (SPD) at telecom-band. Three features of SPDs mainly influence their performance: quantum efficiency (QE), timing jitter and noise.

The most commercial detector at telecom band is an InGaAs/InP avalanched photo diode (APD), where a photo-excited carrier grows into a macroscopic current output via the carrier avalanche multiplication in the APD. The InGaAs/InP APD has a fair QE of typically 10%. However fractions of the carriers trapped in the APD can trigger additional avalanche and cause erroneous counts, which is well known as "afterpulsing". Thus the APD must be operated at gated mode with a relatively low clock frequency. Recently, the InGaAs/InP APD-based SPD has been proved to work at a 1 GHz clock rate with the help of self-differential circuit[19]. However efforts are still needed for a higher clock rate. Superconducting single photon detectors have a low dark count, a small timing jitter. Thank to these characteristics, it has been used in 10 GHz clock system, while its quantum efficiency is as small as 1% and can only be operated at a few kelvin[17].

Periodically poled Lithium Niobate (PPLN) waveguide based frequency up-conversion SPDs have been proved to work at 10 GHz with a quantum efficiency of 8%[15]. It first up converts a photon at telecom band into a visible photon through a non-linear effect, called sum frequency generation (SFG) and then detects the visible photon with a Silicon APD. So far, it seems to be the best candidate for high bit rate QKD, however the timing jitter of the up-conversion Si APD will get larger with count rates increasing. For example in Reference 15, the timing jitter at MHz count rates was larger than 1 ns, therefore this type SPD cannot operate at GHz system.

On the other hand, as there is no commercial single photon source at telecom-band yet[1], an attenuated coherent light from a laser is mainly used as the source for QKD. While the standard BB84 protocol with an attenuated coherence light source will be eavesdropped by a sophisticated PNS attack[7,8]. In a PNS attack, an eavesdropper (Eve) first implements a quantum non-demolition (QND) measurement of the photon number in each weak coherent pulse. If the photon number is more than one, Eve will keep one in her quantum memory and send the others to the receiver through her lossless transmission line. After knowing the measurement basis from the public communication, Eve can utilize the same basis to measure her stored photon without causing any bit errors. Therefore several PNS tolerable protocols have been provided, for example, Scarani-Acin-Ribordy-Gisin protocol[20], decoy state BB84 protocol[21-23], Bennett 92 protocol with a strong reference pulse[24,25], the coherence one-way protocol[26] and DPS[5].

Here we adopted the DPS protocol and up-conversion assisted hybrid photon detectors in the experiment to achieve high bit rate QKD.

In the DPS protocol[5], the sender first encodes one photon in a sequential pulse train, randomly sets the relative phase of each pulse in the train as 0 or $\pi$, and then sends it to the receiver through a fiber link. The receiver uses a 1-bit delay Mach-Zehnder interferometer to make the sequential pulses interfere. Each output of the MZI is connected to a single photon detector, respectively. The relative phase of the sequential pulse decides which detector the photon will go. When one of the detectors fires, the receiver can conclude whether the phase is 0 or $\pi$ according to the which-detector information. The receiver records the phase result and the time instance information when he detects a photon. And then the receiver reports the time instance information, with which the sender can tell the value of the relative phase. Therefore the sender and receiver can share the random phase as the quantum key. This protocol is simple but tolerable against all individual attack including the PNS attack. A detailed proof can be found in Reference 9. Here we provide a rough explanation.

In the PNS attack, a QND measurement on two consecutive pulses breaks the coherence of the pulses, thus brings bit errors. Although Eve can reduce the error probability by increasing the number of pulses simultaneously measured by the QND, Eve's information about the key also decreases. Therefore, the information Eve can obtain by a PNS attack is limited by $2\mu$[9], where μ is the average photon number per pulse.

For general individual attacks, we supposed that Eve could implement an optimal measurement attack, where the collision probability for each bit is

$$P_{c0} = 1 - e^2 - \frac{(1-6e)^2}{2}, \tag{1}$$

where *e* was the innocent system error. Then considering a twofold attack composed of a PNS attack and an optimal measurement attack, an upper bound for the collision probability for n-bit will be[9]

$$P_C = (1 - e^2 - \frac{(1-6e)^2}{2})^{n(1-2\mu)}. \tag{2}$$

As the compression factor $\tau$ with privacy amplification was calculated as

$\tau = -\dfrac{\log_2 P_C}{n}$, the secure key rate $R_{SE}$ was reduced from the sifted key rate $R_{SI}$ as

$$R_{SE} = R_{SI}\{\tau + f(e)[e\log_2 e + (1-e)\log_2(1-e)]\}, \qquad (2)$$

where $f(e)$ characterized the performance of the error correction algorithm. When a SPD was operated at a free-running mode, the sift key rate could be written as[15]

$$R_{SI} = \nu\mu T\eta e^{-\nu\mu T\eta t_d/2}, \qquad (3)$$

where $\nu$, $T$, $\eta$, $t_d$ were repetition rate, channel transmission efficiency, detector's quantum efficiency and dead time of the detection system. In order to achieve high bit rate QKD, we need high clock rate $\nu$ and quantum efficiency $\eta$.

Here we utilized periodically poled Lithium Niobate (PPLN) waveguide based up-conversion hybrid photon detectors in the experiment. We mixed the signal photon with strong pumping light at 980 nm in a wavelength division multiplexing (WDM) coupler and sent them to a fibre pigtailed PPLN waveguide for a SFG[25,26]. This device could convert the 1510 nm signal to a 600 nm sum frequency output with an internal conversion efficiency of 99%. The generated 600 nm photon will go through a short pass dispersive prism to eliminate the pumping light and its second harmonic. At last the generated photon will be detected by a HPD.

The HPD[6] incorporated an avalanched diode (AD) into a vacuum tube to receive and amplify the photoelectron from its cathode. In our case, the photon first injected into the GaAsP cathode (ϕ3mm effective area) and generated a photoelectron. The photoelectron was then accelerated by −8,500 V bias and focused onto the ϕ1mm AD, which is 400 V biased. Then, the electron deposited its kinetic energy in the AD and produced thousands of electron hole pairs, which was called an electron bombarded gain. The generated electrons were drifted in the AD, and further multiplied by ten to hundred times by impact ionization. The total gain was the multiplication of these two gains beyond $10^5$. The quantum efficiency of the HPD at 600 nm was about 25% and the dark count was 1000 Hz. The capacity of the AD was 3.4 pF, so that the rising and falling times were both less than 400 ps.

The up-conversion detector's overall quantum efficiency and noise could be controlled by adjusting the pumping laser of the up-conversion detector as was shown in Figure 1 (a). The noise counts were mainly from the parasitic nonlinear processes in the waveguides and the input fiber[25,26]. We also measured the timing jitters of the up-conversion detectors at high-count rates with a 10 ps pulse input, as shown in Figure 1 (b). From the graph, we could tell that the full width at tenth maximum (FWTM) was less than 200 ps even at 10 MHz count rates for both detectors.

The experimental setup was as shown in Figure 2. A CW laser was first modulated into 2 GHz pulse train by a $LiNbO_3$ intensity modulators (IM), with a pulse duration of 70-ps full width at half maximum (FWHM), 500-ps time interval and a central wavelength of 1550 nm. The IM was driven by a 14 GHz pulse pattern generator (PPG). A 3 GHz pulse pattern generator drove a pseudo random number sequence into a phase modulator to encode the pulse train.

We attenuated the encoded pulse train by around 80 dB and set the average photon number per pulse as 0.2, which is the optimal value for the secure key generation rate[15]. Then we sent the encoded, attenuated pulse through a 10-km dispersion shifted fibre (DSF), which has zero dispersion at 1550-nm band so that the chromatic dispersion induced pulse broadening was negligible compared to the detector's timing jitter, to the receiver. The receiver used a 1 GHz, 1-bit delay Mach Zehnder interferometer[27] (MZI) to interfere the sequential pulse. Each out put of the MZI is directly connected with an up-conversion assisted HPD, respectively. When detector 0 fires, the receiver will know the phase between the two interfered pulses is 0. Similarly, when detector 1 fires, the phase will be π. Therefore whenever the receiver get a detection instance, he and the sender can share a random bit.

The output signals of the up-conversion HPD were first sent to a GHz amplifier and a ps timing discriminator for amplification and pulse shaping, and then went to the time interval analyser (TIA) to record the detection time instances and which-detector information. Therefore, the receiver achieved his random key.

In our experiment, to achieve Mega-bit secure key over 10 km fiber, we set a quantum efficiencies of the detectors as 4% and the noise count rates as 30 kHz with 120 mW pump power. We chose a time window of 280 ps and the noise count rate per window was $8.4 \times 10^{-6}$. Since the timing jitter (FWTM) of the up-conversion assisted HPD, shown in Fig. 1(b), was less than 200 ps even at 10 MHz count rate, the quantum efficiency reduced to 75% at most and there was no error due to tail. The baseline error of the system is about 1% due to the imperfection of the MZI.

In the experiment, sifted keys were actually generated between the sender and receiver, and the error rate was measured by directly comparing the sender's key with the receiver's. For each data point described below, we undertook five runs of QKD sessions, and the error rates and sifted key rates were the averages of the five runs, as was shown in Figure 3. The secure key rates were calculated by putting the experimentally obtained error rates and sifted key rates in equation (3) using the experimental results for the sifted key generation rates and bit error rates. The sender and receiver were located in the same room and we performed 10-km fiber transmission experiment using fiber spools, while other additional data were taken with an optical attenuator simulating fiber loss.

Figure 4 showed the theoretical curves and experimental results for the sifted and secure key generation rate as a function of fiber length. In the figure, the secure key rate with a 10 km fibre was 1.34 M bits per second and the error rate was 1.5%, 1% due to the MZI's imperfection and 0.5% due to the noise count. We observed that the theoretical curves fitted very well with the experimental data except for the last data point, whose error rate was already nearby the security threshold 4.1%.

Our system was also secure against the sequential USD attacks because this type attack do not have any advantage over the individual attacks at lower channel loss condition[10,11,17]. The basic idea of USD attack is that Eve has a local oscillator phase-locked to the coherent laser source of the sender and with a probability of $1-\exp(-2\mu)$, she can unambiguously measure the phase between the sequential pulses. Till now, the most powerful USD attack is a sequential attack with the intensity modulation[11]. With our experimental parameter, we can calculate that all the data points we had taken are all secure against this type attack, except for the data point with the distance of 35 km. Furthermore, according to the experimental parameter, the compression factors in the USD attack with the intensity modulation is bigger than those in the optimal individual attack, which means the optimal individual attack provide the tightest bound in our experiment.

The above results showed that the quantum cryptography system we implemented achieved Megabits secure key rate over 10 km fibre link, which was tolerable to all individual attacks including the PNS attack and all known USD attacks. We hope it would be operable in a standard telecommunication network soon. To extend the Megabits system to longer distance, we could reduce the noise counts caused by the parasitic nonlinear process by using an 1810 nm pump[28]. To achieve higher bit rate, we could either develop special cathode for the SFG wavelength, 600 nm in our case, to 50%, or reduce the cathode thickness to improve the timing jitter.

**Acknowledgements**

Q.Z. thank C. Langrock for his useful discussion during this research. Financial support was provided by the the National Institute of Information and Communications Technology (NICT) of Japan, CREST and SORST programs of the Japan Science and Technology Agency (JST), Hamamatsu Photonics.

**Competing interests statement**

The authors declare that they have no competing financial interests.

**Figure Captions:**

**Figure 1** Performance of an up-conversion assisted HPD. **(a)** Quantum efficiency and noise count rates as a function of pump power. **(b)** Timing jitter as a function of different count rates.

**Figure 2** A schematic drawing of the experiment set-up. The 3 GHz pulse pattern generator (PPG) has two outputs, pseudo random number from data output and 3 MHz synchronized signal from synchronization output. The data output is connected to the phase modulator (PM), while the synchronized signal is used as the start signal of the time interval analyser (TIA). To achieve the perfect phase modulation, a delay line was used to adjust the delay between the laser pulse and the electric pulse of the random number in the PM. The 1 GHz amplifier and timing discriminator were used to amplify the output signals of the up conversion HPDs and formulate standard NIM signal as the inputs to the logic gate. The logical OR signal of the two detectors is used as the stop signal of the TIA.

**Figure 3** Sift key rate (a) and bit error rate (b) of our system as a function of the transmission distance. The 10-km data points were achieved with experimental fiber transmission for both graphs, while other points were with attenuation simulated fiber loss. The curves corresponded to the theoretical predictions with the experimental parameters we set.

**Figure 4** Secure key rate of our system as a function of the transmission distance. The 10-km data points were achieved with experimental fiber transmission. The curves corresponded to the theoretical predictions with the experimental parameters we set, while the data point was calculated from the experimentally measured sift key rate and bit error rate.

Figure 1

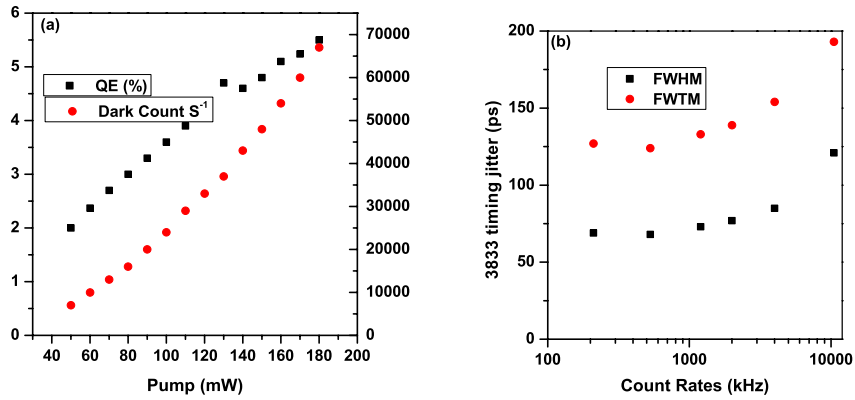

Figure 2

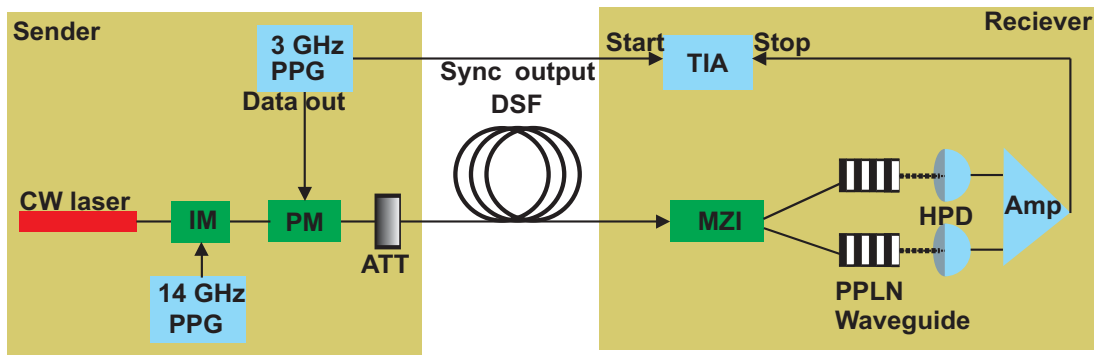

Figure 3

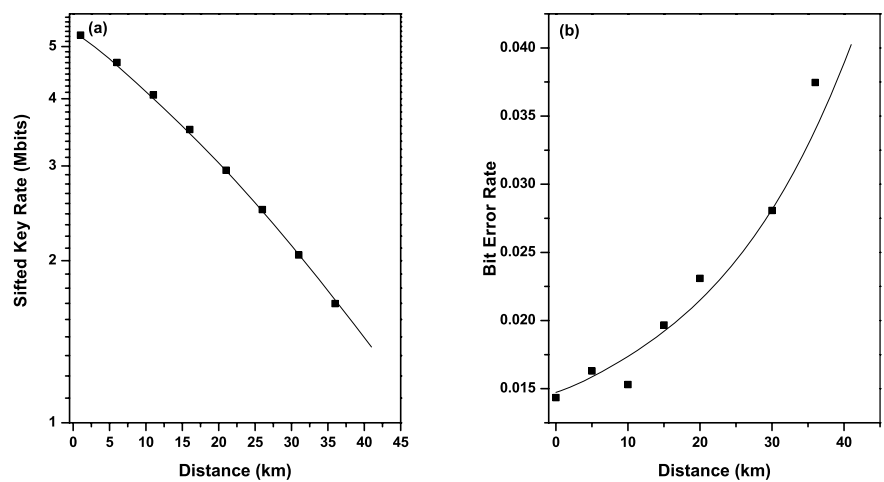

Figure 4

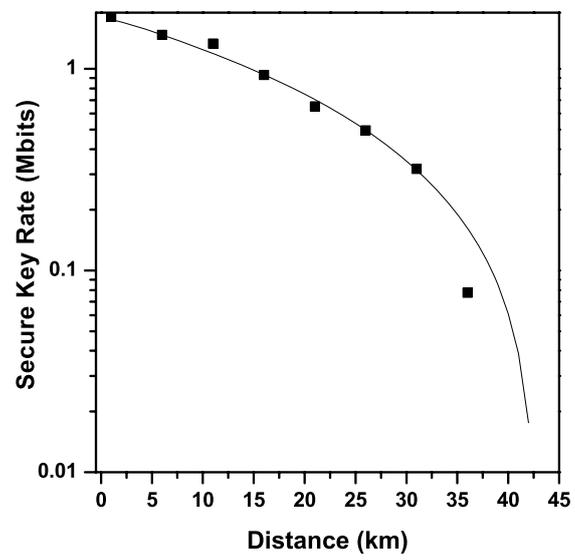